\documentclass[sigconf]{acmart}

\AtBeginDocument{%
  \providecommand\BibTeX{{%
    \normalfont B\kern-0.5em{\scshape i\kern-0.25em b}\kern-0.8em\TeX}}}

\usepackage{graphicx} 
\usepackage{subfig}
\usepackage{hyperref}
\usepackage{balance}

\setcopyright{acmcopyright}

\copyrightyear{2021}
\acmYear{2021}
\setcopyright{acmcopyright}\acmConference[FAccT '21]{Conference on Fairness, Accountability, and Transparency}{March 3--10, 2021}{Virtual Event, Canada}
\acmBooktitle{Conference on Fairness, Accountability, and Transparency (FAccT '21), March 3--10, 2021, Virtual Event, Canada}
\acmPrice{15.00}
\acmDOI{10.1145/3442188.3445897}
\acmISBN{978-1-4503-8309-7/21/03}

\begin{document}

\title{Narratives and Counternarratives on Data Sharing in Africa}
\author{Rediet Abebe}
\affiliation{%
  \institution{University of California, Berkeley}}

\author{Kehinde Aruleba}
\affiliation{%
  \institution{University of the Witwatersrand}}

\author{Abeba Birhane}
\affiliation{%
  \institution{University College Dublin \& Lero}}

\author{Sara Kingsley}
\affiliation{%
  \institution{Carnegie Mellon University}}

\author{George Obaido}
\affiliation{%
  \institution{University of the Witwatersrand}}

\author{Sekou L. Remy}
\affiliation{%
  \institution{IBM Research - Africa}}

\author{Swathi Sadagopan}
\affiliation{%
  \institution{Deloitte}}

\renewcommand{\shortauthors}{}

\begin{abstract}

As machine learning and data science applications grow ever more prevalent, there is an increased focus on data sharing and open data initiatives, particularly in the context of the African continent. Many argue that data sharing can support research and policy design to alleviate poverty, inequality, and derivative effects in Africa. Despite the fact that the datasets in question are often extracted from African communities, conversations around the challenges of accessing and sharing African data are too often driven by non-African stakeholders. These perspectives frequently employ a \emph{deficit narratives}, often focusing on lack of education, training, and technological resources in the continent as the leading causes of friction in the data ecosystem.

We argue that these narratives obfuscate and distort the full complexity of the African data sharing landscape. In particular, we use \emph{storytelling} via fictional personas built from a series of interviews with African data experts to complicate dominant narratives and to provide counternarratives. Coupling these personas with research on data practices within the continent, we identify recurring barriers to data sharing as well as inequities in the distribution of data sharing benefits. In particular, we discuss issues arising from power imbalances resulting from the legacies of colonialism, ethno-centrism, and slavery, disinvestment in building trust, lack of acknowledgement of historical and present-day extractive practices, and Western-centric policies that are ill-suited to the African context. After outlining these problems, we discuss avenues for addressing them when sharing data generated in the continent.
\end{abstract}
\sloppypar


\maketitle

\section{Introduction}
Data sharing is often regarded as a crucial element of research and scientific advancements, a hallmark of transparency, and a key to economic growth \cite{anane2018you, batuo2018linkages, bezuidenhout2018hidden}. Data sharing expands knowledge and contributes to a better science \cite{chawinga2019global, lemarchand2007consociationalism, sebake2013assessing}. Recent years have seen increased global momentum for improving data sharing and, in some cases, a push for open data and lowered barries to data access \cite{bezuidenhout2019share, mwelwa2020developing, academy2019mapping, zaba2013effect, afful2017open, onyancha2016open}.\footnote{While data sharing often focuses on facilitating ease of data transfers between individual scientists, institutes, and/or organizations, truly open data goes beyond that by making data `freely' available for use to a wider-range of populations, rendering it a public good. In this work, we primarily focus on data sharing practices, although many of the insights we draw also apply to arguments for opening data.} However, concerns such as privacy, intellectual property rights, organizational and structural challenges, cultural and social contexts, unjust historical pasts, and potential harms to marginalized communities all make improving data access a highly complex and at times contentious issue \cite{afful2017open, bezuidenhout2016hinders, kaye2012tension, makulilo2016context}.

Data sharing challenges are further exacerbated in the African context: Historical and present-day colonial-era practices and ideologies force us to confront a reality characterized by \emph{data colonialism} \cite{birhane2020algorithmic, coleman2018digital,mkono2019neo,taiwo2010colonialism}. Countless data sharing initiatives as well as a growing enthusiasm for data sharing movements can be found throughout the 54 distinct nations in the continent \cite{anane2018you, batuo2018linkages, chawinga2019global, tchamyou2017information}. At the same time, various obstacles limit equitable data sharing practices; entire heterogeneous geographies of people have their data accessed and shared, yet do not reap the same benefits as the data collectors and owners of data infrastructures \cite{mwaka2017stakeholder, denny2015developing}. These communities can face harm as a result of their data being accessed, shared, and used \cite{munung2016obtaining,nyangulu2019compensation}. At its core, various structural and relational challenges plague the relationships between entities based outside of the continent and the African communities whose data is being accessed. The continent’s plural and at times divergent norms, practices, and traditions furthermore complicate the African data access and sharing ecosystem. In sum, data sharing in Africa poses a unique set of challenges that remain understudied.

\begin{figure}[!ht]
    \centering  
    \includegraphics[width=65mm]{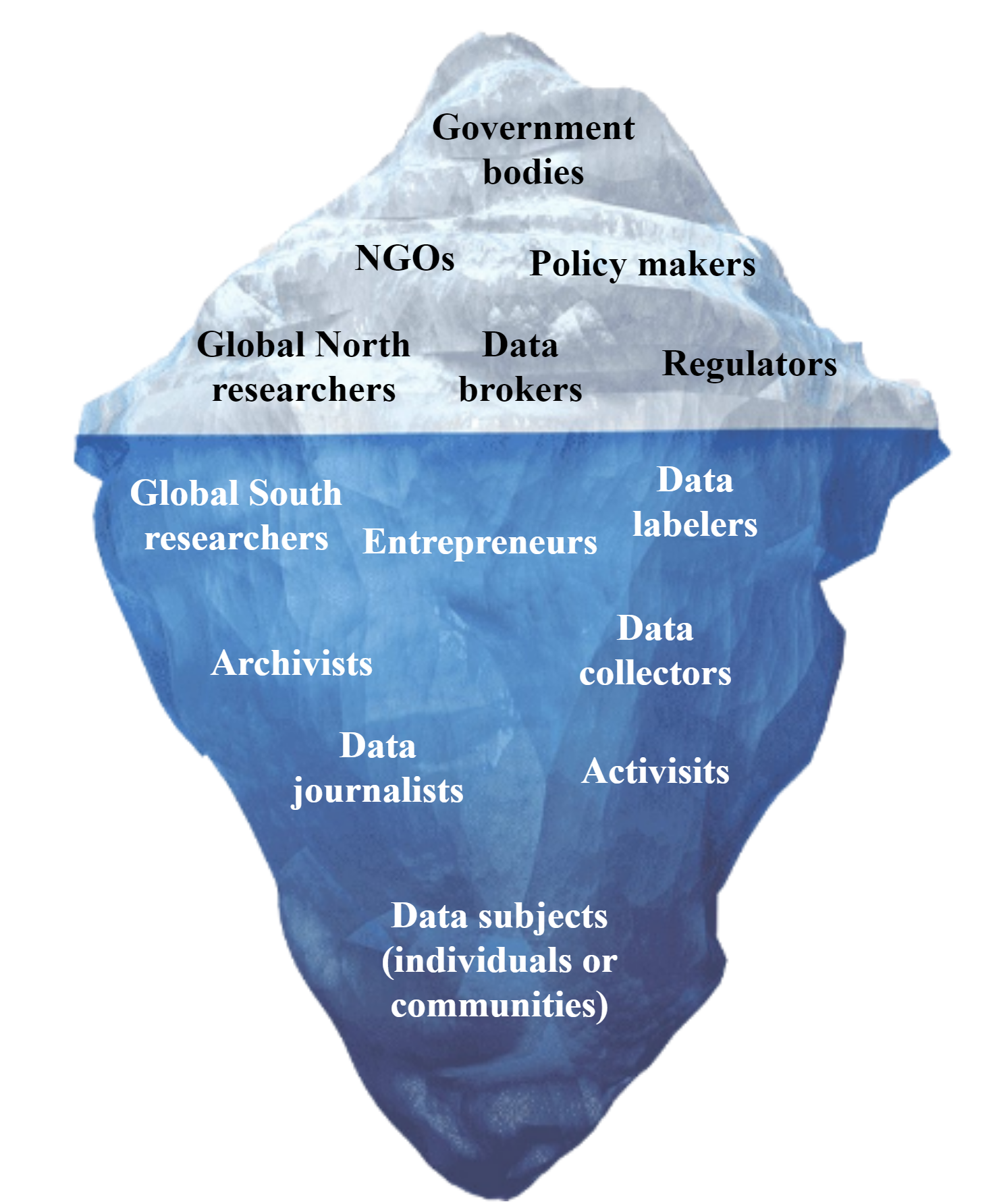}
    \caption{Stakeholders in the African data sharing ecosystem. Those at the top of the iceberg hold significant power and leverage in guiding data sharing practices and policy compared to those in the hidden part of the iceberg. (Image adapted from \cite{d2020data})}\label{fig1aa}
\end{figure}

The African data sharing ecosystem involves numerous stakeholders, including communities, companies, government bodies, policymakers, non-government organizations (NGOs), data collectors, data labellers, data brokers, data subjects, archivists, researchers, and many others \cite{adekunle2013agricultural, mwaka2017stakeholder, musesengwa2018framework, swemmer2015toward}. Responsible data sharing requires a harmonious partnership  between all stakeholders, both under and on top of the iceberg, as depicted in Figure~\ref{fig1aa}. Consideration of the interests, values, and needs of all --- and especially those whose data is being collected --- is a vital step in fostering responsible data sharing practices. In many cases, however, we find that more powerful stakeholders wield disproportionate power in driving the design, framing, and enforcement of data sharing policies and practices; their values and interests dictate not only the data sharing landscape, but also narratives around data access.

On the other hand, data sharing in the continent relies heavily on those in the hidden part of the iceberg. Data is often extracted from these same communities, and analyses of the data often have a direct impact on them. Given the significant role of power asymmetries between these groups, data sharing policies and practices that do not acknowledge such imbalances will fail to capture all relevant perspectives. In this work, we examine this dynamic data sharing landscape, its possible benefits and harms, and challenges around data practices. Our work focuses on the following questions: Who benefits from data sharing in Africa? What barriers exist in the data sharing ecosystem, and for whom? How do entrenched legacies of power disparity, colonialisms, ethno-centricisms, slavery, and Anglo- and Euro-centrism present challenges in the data sharing ecosystem? 
If much of the data sharing practice and initiative is shaped from the perspective of the Global North, how then can we ensure that the narrative for Africa is controlled by Africans? We tackle these questions guided by the experiences and perspectives of African communities and initiatives, and especially those in the hidden part of the iceberg.

Dominant narratives around data sharing in Africa often focus on a lack of knowledge about the value of data and training, as well as insufficient technological resources. This is often followed by \emph{deficit narratives},\footnote{Deficit narratives reduce a group or culture to its ``problems,'' rather than portraying it with the strengths, creativity, and agency of its people.} where discussions regarding data sharing in Africa revolve around issues of hunger, illiteracy, unemployment, and poverty. In anchoring narratives around these concerns --- which are not unique to Africa and do not capture the full scope of its people’s experiences --- the strength, agency, and scientific and cultural contributions of communities within the continent are minimized. Community norms, values, and traditions are overlooked, and policies and practice around improving data sharing focus on ``fixing'' the continent's problems. The perception of the dominant (often Western) narratives as the standard comes at the expense of discarding local knowledge and expertise. Furthermore, data sharing issues are discussed as a Global North/Global South concern, ignoring plural norms, practices, and traditions that exist within the continent, which themselves demand a thorough examination.

As we navigate this data sharing landscape, we counter these deficit narratives and clich\'ed stereotypes with which the African continent is frequently portrayed. In doing so, we hope to highlight overlooked imbalances, inequalities, and injustices in the data sharing ecosystem, and to support the development of policies and programs that will illuminate and mitigate them. We counter these narratives through \emph{storytelling}. We construct fictional personas based on semi-structured interviews with data experts from the continent, as well as our collective expertise in collecting, accessing, and sharing data within the continent. Through these personas, we highlight challenges faced by African communities in accessing and sharing data, as well as the strategies they employ to insist on community-affirming data practices. 

The stories also examine the power structures that affect if and how a community benefits from transactions involving the use of their data. By using these personas as touchpoints, we find that three overarching challenges --- power imbalances among stakeholders, disinvestment on building trust, and disregard for context and local knowledge --- drive inequalities and injustices in the data sharing ecosystem. Through an analysis of these failures, we argue that  the policy and practice of data access and sharing must centre the needs of those stakeholders in the hidden part of the iceberg. The values and expertise of communities whose data is being collected and shared need to take precedence in the development of data governance standards. Through our discussions, we advocate for policies and practices that acknowledge and aim to counter power asymmetries and structural obstacles. In presenting our arguments, we also bring to the forefront rich research and scholarship on this topic by African scholars.

\section{Background and Related Work}\label{Sec:Background}

Data sharing and open data movements are gaining momentum globally \cite{knoppers2011towards, bezuidenhout2018hidden, chawinga2019global,onyancha2016open}, including in the context of the African continent \cite{bezuidenhout2019share, anane2020ethical, chiware2020open,chiware2020open}. Some of this push within the continent comes from funding agencies and government requests for researchers to archive and share data, including open data portals from countries such as Kenya, Ghana, Nigeria, and South Africa \cite{dinah2019role, iglesias2019open,bello2016open,abubakar2019towards}. The call for data sharing is found in various researching fields across the continent, including in the environmental, geospatial, agriculture, health, and biomedical sciences \cite{sebake2012results, denny2015developing, bangani2019data,schaepman2007bridging}. The intergovernmental Group on Earth Observations (GEO)~\cite{AfriGEOSS1}, for example, runs initiatives focused on Africa (AfriGEOSS) with the aim of enhancing earth observation data production, management, and sharing. Similarly, the digitization and sharing of soil data are believed to play a crucial role in a number of fields, including food security, health, hydrological modelling, and climate change \cite{van2019digital, janion2016soil,tesfaye2019actor,gebru2021contributions}. Citing dialogues during a South African Soil Information Workshop, scholars identify the expansion of soil databases and data sharing as important developments for soil science and environmental fields \cite{paterson2015spatial,belete2020partitioning}. The biomedical sciences also echo the call for data sharing \cite{adebamowo2018implementation, glennon2020misinterpretation,bezuidenhout2017beyond,denny2015developing,mutuku2014open}. In a recent paper, \citet{martin2018critical} have called for global genomic data sharing and discussed the urgent need for more genetic data studies, while also arguing that African populations are not represented in global genomic databases. Although the authors raise crucial concerns, including the unsuitability of the ``European method'' to study heterogeneous African populations, they contend that data sharing is vital for a fair representation.

One of the strongest arguments for rethinking how data is shared comes from the Academy of Science of South Africa, which created the African Open Science Platform, a project led by the South African Department of Science and Technology \cite{academy2018african}. This project emphasizes the need to increase awareness, accessibility, and visibility of African science and data. It calls for data and science to be as open as possible, with the aim of stimulating interdisciplinary use of data, identifying research areas for further exploration, and averting the duplication of research.

The questions of data sharing and management are becoming ever more pertinent to the continent. In one of the key gatherings in 2018, various stakeholders and government officials from Botswana, Ethiopia, Kenya, Madagascar, South Africa, and Uganda discussed and formulated national policies \cite{veldsman2018data}. With increased demand for statistical data across many African countries, many data organisations have been established to promote data accessibility and open data \cite{zaba2013effect, afful2017open, bezuidenhout2017technology, chiware2020open, onyancha2016open}. For instance, the African Development Bank Group developed a statistical open data portal known as Africa Information Highway (AIH) portal \cite{OpenData}. AIH links all African countries with the aim of increasing public access to statistics across the continent, while also helping these countries improve data management, data quality, and sharing/dissemination. Similarly, Kenya Open Data \cite{mutuku2014open} is a portal that provides public access to several governmental datasets curated in different categories at no cost. This platform aims to support the government's commitment to transparency and data sharing between government offices and various stakeholders.

Despite extensive calls for data sharing and open data, backed by various initiatives and organizations, data access remains relatively limited within the continent \cite{staunton2019ethical, chiware2020open,dlodlo2012state,tessema2019sustainable}. \citet{bangani2019data}, for example, compare data sharing practices among researchers in South Africa to researchers in countries such as France, the United Kingdom, and Turkey. The authors state that most researchers in South Africa are not interested in sharing their data and prefer to use data generated by others; only 19.4\% of South African researchers indicated that they currently share their research data with others. As reasons for their reluctance, these South African researchers cited worries ranging from unethical use or misuse to lack of resources. In extended fieldwork with biochemistry laboratories in sub-Saharan Africa, \citet{bezuidenhout2016hinders} find that scientists opt for closure over openness with regard to their data. The reasons put forward include inequalities in research environments which, if not examined critically, perpetuate disparities. Furthermore, an overwhelming number of researchers in Africa lack the funding necessary to make their data open to the wider community \cite{bezuidenhout2018hidden}. \citet{bezuidenhout2018hidden} reporte a lack of financial investment in physical infrastructures could make a researcher’s data unavailable, especially for critical clinical research. \citet{barnes2019equitable} agree that funding is a panacea for doing quality research and promoting data openness in low-resource settings. We note, however, that enthusiasm levels for sharing data vary from field to field \cite{serwadda2018open,dlodlo2012state,tessema2019sustainable, anane2018you}.

Obstacles to data sharing are not only financial or infrastructural, but also extend into the political sphere.  \citet{bezuidenhout2016hinders} raise a fundamental question regarding open data and open science. Their study shows that data sharing and open data practices are informed by Western perspectives, and are driven by the contextual issues of Western researchers. Given that different circumstances will give rise to different issues, they go on to ask whether the value attached to data sharing is a specifically Western one. The question becomes: Should the African continent aspire to develop its own data sharing policies and initiatives grounded in its distinct values, context, and communal culture? This perspective takes data sharing challenges beyond the financial and technological realms and suggests the need for  a fundamental rethinking. \citet{yilma2013spatially}, for example, has proposed the absence of a strong political will as a major barrier to data sharing in Africa. Likening data sharing practices to aid programs, Yilma states, ``any type of aid program that does not lead eventually to self-sufficiency is actually destructive, just like welfare, and that is what has happened with many aid programs in Africa'' \cite{yilma2013spatially}.

\section{Storytelling as a Methodology}

\begin{quote}
  \emph{``Stories are just data with soul"} -- \citet{brene2020}.
\end{quote}

In this section, we illustrate the importance and relevance of storytelling to our work. We first trace its historical significance to the African continent before discussing our methodology for creating personas aimed at presenting counternarratives.  


\subsection{Storytelling in Africa}
Oral culture and the African continent are profoundly linked in most spheres, from poetry and dance to philosophical traditions~\cite{n1971literary,olaoluwa2008inscribing,soyinka2020dionysus,tuwe2016african}. Many philosophical ideas have deep roots in the African continent \cite{henshilwood2002emergence, mcdougall2005stratigraphic,rananga2008professionalising,mojapelo2020storytelling}.
Ancient Egypt, for example,  marks the oldest known philosophical tradition~\cite{ohaegbulam1990towards,kurewa2016african,ilahiane2017historical}. Furthermore, in pre-colonial Africa philosophical practices among indigenous peoples took the form of oral traditions~\cite{seroto2011indigenous,imbo2002oral,adeyemi2003principles}. Emphasizing the distinctness of oral traditions to the African continent, Okot p’Bitek \cite{p1975fr} contends that oral practices are much more immediate and powerful. In \textit{Oral Traditions as Philosophy: Okot p’Bitek’s Legacy For African Philosophy}, \citet{imbo2002oral} further demonstrates the unique significance of oral traditions in African philosophy.

Storytelling is one important method for indigenous research, and a tool to legally claim the right to local voices through which narratives are validated \cite{delgado1989storytelling, gachago2014all,mataira2019transforming,inamara2017pacific}. Making local communities the focal perspective of shared knowledge through storytelling counters histories of colonialism, knowledge produced for colonial regimes, and the power dynamics silencing indigenous expertise \cite{agan2006storytelling, merolla2019memory,shizha2014indigenous,ayodele2020colonialism}. In South Africa, for example, the \textit{Truth and Reconciliation Commission} (TRC) incorporated storytelling -- songs, poems, and oral literature -- into the official legal process of speaking aloud marginalized experiences of Apartheid \cite{oboe2007trc}. In Uganda, the National Education Ministry mandates storytelling as part of the required national curriculum. The Ugandan Education Ministry's curriculum reform since independence has likewise re-introduced storytelling by law. These reforms are focused on re-instituting local knowledge and teaching methods, as well as countering the influences of former colonial powers \cite{ssentanda2019reclaiming,anyeko2020storytelling,mudondopast,ngaka2020role}. The African StoryBook Project\footnote{\url{https://www.africanstorybook.org/}}, a repository of stories in local African languages, illustrates the value of both storytelling as a method of knowledge creation and data sharing practices driven by community needs. This open source repository of digital books written in local African languages allows the re-purposing, scaling, and sharing of content from the platform, helping the Ugandan Education system to make the switch to teaching in local languages\footnote{Although the African StoryBook Project can be seen as a powerful example, it is also important to acknowledge that limitations and challenges remain for the Ugandan Education system. For instance, the harms of colonial regimes are not extinguished immediately upon instituting policy reforms. The dynamics of the global political economy also continue to pressure national education systems to emphasize teaching and establishing grading or progress standards for students that are measured by English language competency, e.g. standardized tests}.

African researchers and scientists have called for the international scientific community to embrace indigenous research \cite{corntassel2009indigenous, baskin2005storytelling,chilisa2019indigenous}. The call acknowledges how research design (e.g. how ``validity'' is determined) could reflect Western values, and thus, co-opt, erase, and/or circumvent legitimate means of expressing knowledge. Problematically, many Global North research methods characterize local ways of creating and validating knowledge as \textit{scientifically invalid} \cite{dyll2014development,wuni2019scientometric,kiraka2020university}. Computational journal and conference reviewers also insist on Global South authors contextualizing research, including methods, in relation to the Global North. We argue this demand erases and marginalizes local expertise.

\subsection{Storytelling of African Data}
Inspired by the role of storytelling in the African context, we illuminate challenges faced when accessing and sharing data in the African continent from African perspectives. We do so through personas created to portray recurring challenges we encountered throughout our interviews, as well as our experiences of collecting, accessing, sharing, and using data generated in the continent.

We constructed the personas through an iterative process based on semi-structured interviews with ten African data experts. Our interviewees included individuals working in universities, governments, startups, and think tanks, and we focused on data collected and shared for research or policy purposes. We first developed open-ended questions, guided by our experiences observing and actuating data sharing practices on the continent. The interviews were recorded with permission from the interviewees, after which we manually transcribed the audio recordings. Next, we identified and abstracted central themes, challenges, and concerns expressed by the interviewees. Any information that could be used to identify our interviewees, institutions, government bodies, or communities was removed. In addition, we only took a particular  challenge as the inspiration for the creation of a persona if that challenge  appeared in at least three interviews. These decisions allowed us to preserve the privacy of both the individuals we interviewed and the communities of concern, while still creating space for discussing common data sharing practices that are not affirming of the needs, values, and concerns of those on the bottom of the iceberg. In response to these privacy concerns, we did not incorporate any direct quotations from our interviews into the personas, but we plan to include data of this kind in subsequent publications. 

Each persona consists of two to three core challenges. Over the course of multiple meetings, we discussed and debated the efficacy of each persona in portraying the challenges we would like to elucidate. In our discussions, we primarily focused on the following key questions:

\begin{enumerate}
    \item Do the personas elucidate the intended data sharing challenges, and if not, why not? How can we better bring these challenges to the surface?  
    \item Which data sharing challenges are more ``familiar'' to a diverse audience, including many researchers from the Global North? In what ways should we amend the personas to hone in on what challenges may be unique to the continent?
\end{enumerate}

This process allowed the personas to evolve and solidify over time. We rebalanced the emphasis on various themes in the personas at times by adding details from more interviews or removing details from interviews that obfuscated the key challenges we would like to highlight. In this paper, we present a sample of three personas that illustrate the three key overarching challenges we found in our work. Additional personas and details about this project are shared on our project page.\footnote{\url{https://www.md4sg.com/workinggroups/GlobalSouthData.html}.}

\section{Personas}

\begin{quote}
   \textit{``Proverbs are horses on which words ride, when words are lost [i.e. when the truth of a matter is elusive], it is proverbs we use to search for them [i.e. proverbs help us to solve an intractable problem]''} -- Agbájé~\cite{agbaje2002proverbs}.
\end{quote}

The personas we present below present recurring themes we encountered in our interviews and subsequent discussions. These stories are not meant to reflect the experiences of any single individual, community, company, or country, but rather to gather evidence and shed light on recurring barriers. Since these stories are based on aggregate information, they call attention to common social issues while respecting sensitive dynamics -- such as concerns about privacy and retaliation -- that may make interviewees, and indeed many across the continent, vulnerable when sharing these experiences. As such, these stories convey and conceal truths at the same time.  

\vspace{3mm}

\noindent \textbf{On Good and Harm}\newline
\textit{Themes: power imbalance, data colonialism }

\vspace{2mm}
\noindent \underline{Context:}\newline
DrinkUS West is a Europe-based non-government organization (NGO) working on sustainable access to water and health. DrinkUS hopes to support people in Buranda -- home to the yetet'ebek'e community -- by improving access to clean, potable water. It also plans to deploy and test new water accessibility technology and online monitoring of resources.
\medskip

\noindent \underline{Data needs:}\newline
To provide adequate services and assess its impact, DrinkUS seeks data about people's day-to-day water use habits and experiences through surveys. In the survey, DrinkUS researchers ask individual respondents about their demographics (age, gender, income, marital status, number of household members and children), discretionary spending habits (ownership of computers, mobile phones, leisurely activities), and water use habits (such as how water and utilities are accessed, and the respondent's knowledge of potable water accessibility). Additionally, DrinkUS seeks country-wide data on demographic patterns as well as outcomes related to health, education, and poverty levels over the past decade. It gains access to this data from the Buranda Health Ministry.

By using these two data sources, DrinkUS aims to estimate demand for potable water, determine prices, and predict the impact such a service may have on the health and economic welfare of the community. Additionally, the NGO assesses technologies that could be used to monitor and manage water access. As the NGO is keen to engage survey participants in future opportunities, it collects contact information such as the names, addresses, and phone numbers of participants. DrinkUS incentivizes each respondent by paying the equivalent of \$5 U.S. dollars in Kwacha, the local currency, for completing the estimated 30-minute long survey to improve survey completion rate. From this data, DrinkUS researchers compile and infer which members lack access to potable water, what community members can afford to pay, and who is most at risk for water-related health impacts and/or disease. They combine these results with data obtained from the Buranda Health Ministry to gain deeper insights into the impact of access to potable water. This work is later published as a research paper at a top-tier international conference on data science for social good.

\medskip

\noindent \underline{Data sharing challenges:}
\newline
While employees at DrinkUS meticulously checked their survey design, they did not seek adequate feedback from the yetet'ebek'e community beyond a few initial meetings. The yetet'ebek'e community is not treated as a mutual partner in the broader project design. In fact, there is an implicit assumption that improving water access for the community is an adequate benefit, and that DrinkUS is therefore justified in collecting and owning the survey data. Further, since DrinkUS is a world-renowned NGO working in partnership with the Burandan government, it wields significant power over members of the yetet'ebek'e community. Even the pay incentives, which may be negligible for the NGO, are significant due to currency exchange rates, further highlighting a power imbalance.

Given that the NGO proceeded without consulting the community, it likely failed to grasp all of the central issues and potential harms from the perspective of the community. Sharing their research findings publicly, including their predictions about which geographic locations in the yetet’ebek’e community were at risk for water-related disease, put the community at risk, as many belong to marginalized ethnic groups. Such findings, even shared in aggregate, can reveal information on the community as a whole. While notions of privacy often focus on the individual \cite{horvitz2015data, kostkova2016owns}, there is growing awareness that collective identity is also important within many African communities, and that sharing aggregate information about communities can also be regarded as a privacy violation \cite{enyew2016towards,namara2018cross}.

In addition, given the partnership with the Burandan government and poorly defined protocols around data sharing and use, survey participant data was shared with government officials, including data about household composition and reported or predicted health challenges linked to potable water access. While the researchers vetted their questions for survey design validity, they did not seek active involvement from the yetet’ebek’e community to discuss whom the data can be shared with and for what purposes. DrinkUS also failed to solicit feedback on the broader project design, data ownership and its use beyond original scope of work, and research procedures. This disempowerment of local communities in the decision-making process introduced trust issues both with the NGO as well as with the Burandan government, which has been known to engage in open data sharing practices with large NGOs and institutions. Local communities, on the other hand, cannot access their own data with the same ease.

\vspace{3mm}
\noindent \textbf{The Journey of African Scholars}\newline
\textit{Themes: trust, awareness}

\vspace{2mm}
\noindent \underline{Context:} \newline
A doctoral candidate from Bozatta researching in Nova Africa wants to collect soil samples from several Nova Africa provinces for her research. Her study aims to investigate the fertility of soil samples, which will inform key stakeholders, such as the Government and NGOs, to find the best approaches to assisting farmers.

In another case study, a doctoral candidate is researching the prevalence of gender-based violence in Nova Africa. The study is targeted towards raising awareness of gender-based violence prevalence, which remains a profound and widespread problem in Nova Africa.\\

\noindent \underline{Data needs:}\newline

\noindent\textbf{Case 1: Soil and apartheid}

\noindent The researcher has outlined critical steps to conducting the soil fertility test, using techniques such as sampling, analysis, interpretation, and recommendation. All of these methods are essential for estimating soil fertility. Enthusiastic about undertaking her research, the researcher aims to collect these samples to determine soil fertility, nutrient deficiencies, and recommendations, such as selecting the most favourable fertilizer plan to increase farmers' productivity. The information gathered from the study will be used to create an effective fertility management program and provide findings to the public through her dissertation and institutional public databases.  

\medskip

\noindent\textbf{Case 2: Gender-based violence}\newline
\noindent The researcher aims to collect this data and create a data platform to present gender-based violence cases within provinces in Nova Africa. The tool will aid data visualisation of these cases and will be available to the broader public. The researcher understands that her work is focused on a pressing issue: Gender-based violence remains a persistent problem, requiring attention from multi-faceted responses and commitments from the government, civil society, and other citizens. In particular, the government does not keep track of data on the prevalence of gender-based violence, despite desire to work towards protections for the vulnerable. The researcher's work aims to narrow this data gap.\\

\noindent \underline{Data sharing challenges:}\newline
\noindent In \textit{Case 1}, in an attempt to collect soil samples, the researcher learns that farmers are not enthusiastic about sharing this information, but she is unable to understand why. After months of attempts, she eventually learns from her colleagues that farmers have collected such data for their own internal purposes, and use it to organize within their communities. The researcher attempts to instead obtain this data, assuming that the farmers' reluctance may be partially due to replication of efforts. She eventually learns that the farmers are reluctant to share any information due to fear that the government might want to claim ownership of their lands. Collection of such data is a challenge for local researchers as well. This issue is exacerbated by the history of a former apartheid regime, under whose reign land grabs were widespread, impacting Black Nova Africans primarily. Since sharing data can lead to forfeiture of land rights, trust is a central issue for the farmers in deciding how much information they share and with whom. \\

\noindent In \textit{Case 2}, during the process of conducting the research, the doctoral candidate experiences a few unexpected encounters and shares them with some colleagues from Nova Africa. She tells them that many participants became emotional when sharing their experiences, and asked that the session be discontinued. One participant even questioned why the interview was so intrusive, and decided not to share further information. Her officemate tells her that in Nova Africa, it is challenging to convince participants to release information, especially as a foreigner. There is a perceived notion that foreigners are here to steal from them, whether that be data or resources. This, she learns, is in part reflective of and caused by divisions across communities in Africa due to colonialism. Another colleague advises her that some participants might seek secluded venues that are secure for sharing experiences in most rural areas, but that building trust is crucial, especially as an outsider of this community.\\

Young researchers working across different African countries may encounter challenges when trying to collect, access, and share data. This challenge is particularly stark when the researcher is not a local to the region or community in which the data is being collected. Differences in languages and cultures, as well as  imbalances in trust and power, breed suspicion about motives, impeding data access and sharing. With more than 1500 languages spoken across the African continent, locals residing in rural areas are more likely to communicate in their dialect, with potentially little or no understanding of national languages. This concern may, at times, stand in contrast to the fact that there is a pressing need on high-quality data to understand the prevalence and impact of issues such as gender-based violence, among numerous other forms of injustices and discrimination.

Discussions on data sharing in the African context often treat the continent as a monolith and frame problems as a Global North/Global South issue. However, the complex data sharing ecosystem can lead to challenges even for those based in the continent. We further find that many issues with power imbalances and trust can be traced back to the impact of entities based outside of the continent. We see, for instance, that certain hesitations with sharing data are rooted in colonial-era extracting practices whose impact can be felt to this day. Numerous efforts to mitigate inequalities in data access and sharing in the continent, and especially those driven by initiatives outside of the continent, may not be attuned to the complex data landscape resulting from a diversity of needs, priorities, and experiences of those in the continent.

\vspace{3mm}
\noindent \textbf{Livestocks and Livelihoods }\newline
\textit{Themes: data infrastructure, awareness }

\vspace{2mm}
\noindent \underline{Context:}

\noindent A researcher in Wolonda receives a two-year grant from a large foundation from the Global North. The research project outlines an ambitious goal of ushering in prosperity for the livestock farmers by making the mainstream markets work for them.
\medskip

\noindent \underline{Data needs:}\newline
The researcher outlined his intentions to collect various types of data,including genomic data of crops and livestock and enterprise data for the farm. The genomic dataset would help determine if the right crops were being planted and the right animal breeds being reared to support the production system. It would help provide the farmers vital intelligence on susceptibility of their crops and animals to diseases and performance of their farm against aggregate data. The project was structured as a donor-funded research endeavor, with clear data collection and analysis objectives. Yet no protocols for data storage and scope of use were defined,  and there were no clear guidelines on whether the data can be shared with third parties. Furthermore, scant information was provided as to if and how the collected data would be used beyond the time period of the donor-funded project.

As expected, the researcher and his team diligently collected the genomic data as well as the farm data from a cluster of villages in several counties of Southern Wolonda. The data collection process was, however, more challenging than originally anticipated. Even with a local team well-versed in the norms and practices of the region, farmers seemed skeptical to give up their data. Slowly, as the end of the two-year grant period approached, the data collection exercise also came to a close. \\

\noindent \underline{Data sharing challenges:}\newline
Encouraged by this newly collected and valuable dataset, a research team established a for-profit company using the data collected as part of the research project. In its mission statement, the company articulated the need for precision agriculture in today’s data-rich environment as an enabler of better agricultural practices. The source data on which the company was built came from that dataset which was collected as part of the original research project. In addition, due to infrastructure challenges, the data is now housed in a data center at a University in the Eastern United Provinces, an influential country in the Global North which is also home to the foundation that originally funded this research.

The company has set its sights on expanding across Wolonda. Even though the trust established with the farmers in Southern Wolonda has eroded due to the actions of the researcher, he is confident that other farmers will see the potential of his data engine and believe in the power of its predictive analytics.

Foundations in the Global North may be eager to fund data collection projects in an attempt to create an oasis in regions perceived to be data-starved -- to bridge data gaps and inequalities that appear to be merely inefficiencies. Yet, as is the case in this scenario, the lack of forethought into the ethical use of data -- not only during the lifetime of the project, but also thereafter -- can create  irreparable harms to communities' well-being and make it difficult to build  partnerships based on respect and mutual trust. This above case may alienate the farmers, precisely the people whom this project was ostensibly seeking to support.

\section{Data Sharing Challenges}\label{DSC}

As the personas in the last section show, obstacles, issues, and challenges of data sharing concerning the African continent are multifaceted. Questions such as, ‘Is data sharing beneficial? Valuable?’ need to be contextualized by subcommunity: We need to also ask ‘for whom?’ Throughout this work, we have examined such questions in a manner that is informed by the perspective of the stakeholders at the bottom of the data ecology chain -- the hidden part of the iceberg. There are numerous causes for challenges in this ecosystem. In this section, we focus on three broad and interrelated challenges that surfaced from our interviews and research. In particular, we explore: power asymmetries, issues of trust, and the need for contextual knowledge.

\subsection{Power asymmetries}
Historically, traditional colonial powers sought unilateral domination over colonized people through control of socio-economic affairs and the reinvention of  social orders for their own benefit \cite{okinedo2019impacts}. In the current day-and-age, when data and digital technologies are powerful instruments, colonial-era oppression has been reincarnated in various data practices, including data collection, sharing, and analysis \cite{birhane2020algorithmic, kim2019}. The search for data accumulation, especially with regard to Western technology monopolies — both the scale as well as the manner in which it is being collected — have raised questions regarding unprecedented wealth accumulation and power struggles. Currently, a significant proportion of Africa’s digital infrastructure is controlled by Western technology powers such as Amazon, Google, Facebook, and Uber. Traditional colonial powers pursued colonial invasion through justifications, such as ``educating the uneducated.'' Data accumulation processes are accompanied by similar colonial rhetoric, such as ``liberating the bottom billion,'' ``helping the unbanked,'' ``connecting the unconnected,'' and using data to ``leapfrog poverty.'' However, this rhetoric may not only preserves historical coloniality \cite{mohamed2020decolonial} dressed in data, but also perpetuates deficit narratives. These objections have been articulated by scholars and technologists within the continent. For instance, \citet{kim2019} writes:
\begin{quote}
    ``I find it hard to reconcile a group of American corporations, far removed from the realities of Africans, machinating a grand plan on how to save the unbanked women of Africa. Especially when you consider their recent history of data privacy breaches (Facebook) and worker exploitation (Uber)." \cite{kim2019}
\end{quote}

Colonial legacies and power imbalances embedded in data practices may appear less obvious and more nuanced within scientific research settings \cite{ashcroft2013post}. Within the context of global health research, there has been growing concern about power imbalances in authorship, which has negatively affected how research undertaken by local researchers in low-income and middle-income countries (especially in Africa) is perceived \cite{abimbola2019foreign}. Academics in the Global North are bestowed with the power to define what constitutes ``legitimate knowledge,'' ``good research,'' ``standard method,'' or a relevant and worthwhile problem. As a result, African researchers are left little room, if any at all, to compete at the global stage, even in matters concerning the African continent. Following extensive analysis of interviews with senior university research managers in Zimbabwe, and on a public roundtable on Structural Inequalities in Global Academic Publishing, Jeater~\cite{jeater2018academic} finds that ``When we ask who gets to represent the `African perspective,' we find it is decreasingly un-African.'' Resource inequalities, hegemonic academic standards that undervalue Southern research traditions, and the unilateral power Northerners hold to validate research, all contribute to structural obstacles that amount to systemic exclusion of African scholarship from global health research. In a data ecosystem built on such firm yet invisible power asymmetries, stakeholders already in a position of power not only benefit the most, but also make data accessibility inequitable.

Imbalances in authorship and power asymmetries constitute a continuation of the colonial project in global health research, creating fundamental trust issues concerning data sharing. Underneath power asymmetries in global health research partnerships between researchers from the North and South, lies what Abimbola~\cite{abimbola2019foreign} calls ``the foreign gaze.'' In his analysis of this concept, Abimbola asks questions such as ``who we are as authors, who we imagine we write for (i.e., gaze), and the position or standpoint from which we write (i.e., pose).'' Furthermore, power asymmetries which occupy an important space around trust and data sharing often operate in invisible ways and take many forms.

In a similar study that explores power imbalances between the global North and South, using Zambia as a case study, ~\citet{walsh2016way} report that power imbalances and inequalities manifest at all stages of research. This includes everything from funding to agenda setting, data collection, analysis, interpretation, and reporting of results. Looking at the underfunded Zambian health research, where up to 90\% of the funding for health research comes from external funders \cite{walsh2016way}, the bargaining power rests with the funders with little room for negotiations for Zambian scholars. Power asymmetries are also observed in the personas above. In ``\textit{On Good and Harm}'', we find the European NGO at the top of the power hierarchy making key decisions regarding the data concerning yetet’ebek’e communities. Similarly, in the ``\textit{Livestocks and Livelihoods}'' persona, the foundation from the Global North which provided grants dedicated to research in Wolonda holds much greater authority and power compared to the villagers who are the source of the data.    

Power asymmetries, historically inherited from the colonial era, often get carried over into data practices and manifest themselves in various forms, from imbalanced authorship to uneven bargaining powers that come with funding. Having said that, power asymmetries are not limited to historical contexts only. Within a given research project, for example, one can observe that power asymmetries exist between project managers and data analysts; data analysts and data collectors; data collectors and research participants. All these factors are contingent on various caveats and implications for trust impacting data processes from data quality to data sharing. Examining data production and consumption process in the context of Malawian demographic survey, Biruk~\cite{biruk2018cooking} makes these power asymmetries, hierarchies and structural inequalities visible. Although obscured by partnership rhetoric, Malawian demographics mapping in fact embodies structures with unequal division of labor. Biruk explains:
\begin{description}
\item ``[B]eing on the ground in the field has the largest effect on data but—from the perspective of researchers—the activities of fieldworkers are framed as menial labor performed by easily replaceable and interchangeable individuals. The local expertise they offer, then, is not in designing research or writing proposals but comes as an additive to a project conceived in a distant office. These hierarchies are embedded in political-economic structures that privilege the knowledge work that is the purview of Western academic researchers over the so-called unskilled labor performed by field workers. Meanwhile, Malawian research collaborators occupy a middle space that is both constructed by and fraught with power and economic inequalities.'' \cite{biruk2018cooking}  
\end{description}

\subsection{Trust}

\begin{quote}
\textit{``Data often move at the speed of trust.''} -- Hamilton and Hopkins \cite{goodman}
\end{quote}

Sharing data between different stakeholders hinges on trust. Trust is the fundamental component of all relationships in a data sharing ecosystem. While trust, or lack thereof, has been identified as a key challenge that hampers data sharing \cite{colquitt2007trust, jao2015research,adogame2014reinventing,serwadda2018open}, there remains much to be examined about the role of trust and how it manifests in relationships between various stakeholders in the African data sharing ecosystem.
Data sharing practices which operate in the absence of knowledge of local norms and contexts contribute -- albeit indirectly -- to the erosion of trust among stakeholders in the data sharing ecosystem. Initiatives coming from outside, with their own assumptions, interests, and objectives, tend to be met with suspicion by local communities. The persona on ``\textit{Soil and Apartheid}'' captures this in a stark manner; due to the apartheid regime, the doctoral researcher finds that Black farmers in Nova Africa suffered unjust land grabs. Such historical injustice plays out in the farmers' reluctance to share soil data due to lack of trust.  

Resource inequalities and colonial oppressive histories instill deep mistrust towards open data and data sharing initiatives. Although African researchers are generally supportive of data sharing, they are considerably less enthusiastic about open data expressing concerns that open data compromises national ownership and reopens the gates for ``parachute-research'' (i.e., Global Northern researchers absconding with data to their home countries).  
Such concerns are not unwarranted. In fact, the findings from a recent study from \citet{mbaye2019telling} affirm this fear. \citet{mbaye2019telling} performed a systematic review examining African author proportions in the biomedical literature published between 1980 - 2016, in which research was originally done in Africa. The authors found that African researchers are significantly under-represented in the global health community, even when the data originates from Africa. A common threat is parachute-research, in which non-African researchers benefit from data sharing and open data,  are afforded the opportunity to narrate African stories (in some cases also contributing to deficit narratives), and publish scientific work using African generated data available through open access initiatives -- all while ignoring the contributions of African communities and scholars ~\cite{kraemer2020dynamics,yozwiak2016roots,moroke2015train,mathebane2017towards}. Recent work within the medical sciences published on the Ebola outbreak in the Democratic Republic of the Congo can be cited as a prime example in this regard.

Ideal data sharing initiatives, policies, and principles weigh the benefits and potential risks, and strive to find a reasonable balance. Benefits and risks also vary depending on the types of data being shared. For example, the use of genomics data poses the potential for far more detrimental risk to the individual providing the data, to the researcher, to the institution, and to the community compared to soil data. The issues of trust as a challenge to data sharing, likewise, vary in degree depending on the data in question. Trust is a relatively significant challenge to sharing sensitive biomedical or health data and less significant when it comes to, for example, environmental sciences data. Noting this concern, \citet{walport2011sharing} emphasize, ``people often agree to provide sensitive data because they trust the researcher and believe the researcher would not use the data in a way that would be harmful towards them. However, there is a concern that the trust may not carry over when the data are shared''.

\subsection{Contexts and local knowledge}

Contexts are a crucial element to making sense of data and data sharing. Yet data -- within data science and machine learning, particularly -- are often stripped away of contexts~\cite{magrin2009use,cini2019disrupting,misago2019political}. In the process of data cleaning, for example, information that provides contexts about the specific background from which data are collected and how datasets are structured can be lost~\cite{Aremudata}. The importance of context for datasets has been explored by scholars, such as \citet{loukissas2019all}, according to whom, we should shift into thinking in terms of \textit{data settings} instead of \textit{datasets}. Contexts are crucial to understanding data fully; data sharing practices that discard contexts risk becoming irrelevant and potentially harmful to local communities, as we see in each of the above personas.

Consequently, the context of the data itself -- which provides a complete image -- and awareness of local norms, cultures, and histories constitute crucial elements in a responsible data sharing practice. Thus, for data sharing practices to benefit the underserved, such groups' welfare and interests need to be placed at the center stage. However, our interviewees nearly unanimously agree that there remains a lot to be done to acknowledge and incorporate the interest, norms, and context of these below the iceberg in Figure \ref{fig1aa}. In fact, oftentimes, certain groups such as data subjects are hardly recognized as stakeholders at all~\cite{alaneme2018critical,gillwalddata,keane2017decolonising}.

We see various levels of disregard for context and local norms displayed in all the personas portrayed in the previous section. In "\textit{On Good and Harm}", the failure to ground data collection in communal understanding of privacy, for example, resulted in the NGO exposing yetet'ebek'e community to various risks in the process of data sharing. Lack of common language and understanding of local norms creates a challenge for intra-continental data practice, as shown in \textit{"The Journey of African Scholars."} In both cases, the doctoral candidate faced difficulties understanding the local norms, which played a role in the community's lack of trust towards her. Similarly, in the \textit{"Livestocks and Livelihoods"} persona, we observe that the funding foundation (from the Global North) failed to consider the context in which the data was collected. In summary, data sharing calls that are not aware of local norms, contexts, and culture, when imposed from the outside, constitutes a form of Western-centrism and colonialism.

\section{Discussion}

Returning to the question ``is data sharing good/beneficial?'' we argue that responsible data sharing practices must, first and foremost, benefit local communities and experts, with a focus on those at the bottom of the iceberg. In recent years, the African continent as a whole has been considered a frontier opportunity for building data collection infrastructures. The enthusiasm around data sharing, and especially in machine learning or data science for development/social good settings, has ranged from tempered discussions around new research avenues to proclamations that ``the AI invasion is coming to Africa (and it's a good thing)" \cite{novitske2018ai}. In this work, we echo previous discussions that this can lead to data colonialism and significant, irreparable harm to communities. As we learned from the rich body of previous works, our experiences, interviews and the personas, data sharing practices are divergent, ad hoc, at times contradictory, and/or violate community values for data use. As machine learning and data science move to focus on the Global South and especially the African continent, the need to understand what challenges exist in data sharing, and how we can improve data practices become more pressing.

Despite significant variation among the different countries on the continent, many countries are now producing high-quality data in a more reliable and regular manner through statistical offices, national efforts, and involvement of other public agencies. Likewise, non-public sectors actors --- such as private-sector companies, academic institutions, civil society organisations, development agencies, and individuals or communities --- are now encouraged and incentivized to be involved in open data management.
The future of open data management and data sharing and their contribution to the advancement of science and technology in Africa will continue to increase, despite the slow pace caused by the lack of funding, redundant policy frameworks, and limited infrastructures. The unique African landscape, and especially the existing challenges and how they can be addressed, will continue to play a big part in African participation in open science and open data global projects~\cite{bezuidenhout2018hidden}.

As explained by \citet{bangani2019data}, to understand the sharing practices of researchers, it is essential first to know where the data comes from and the lives, cultures, and communities it represents. \citet{bangani2019data} also conclude that most of the emerging researchers in South Africa would prefer to use data from other sources and are not interested in sharing their own data. Most of these emerging researchers indicated a lack of resources, misuse of data, and trust issues as the significant challenges they face in sharing their data. Some of them indicated that they would prefer to only share their data with collaborators within their universities and not with collaborators outside their universities. For many of the researchers in this context, the labour, effort, and time it takes to collect data is unmatched by the benefits (often little if any at all) they might receive from data sharing. Subsequently, as well as encouraging researchers to share their data, there also needs to be actionable benefits and rewards, as well as policies that protect African researchers and communities from practices like parachute research.  

Open data imposes risks to African communities -- in terms of both stripping data away from contexts and reduced benefits to communities where data is sourced. It may present a possibility for Global North researchers to publish research using data from Global South with neither acknowledgements to data collectors or data subjects nor an understanding of the data setting. Consequently, there should be less pressure to open up data and more incentives to share data with the necessary conditions and support for infrastructures. The necessary conditions could include data sharing guidelines that give adequate time to African researchers to make use of data before wider sharing \cite{tenopir2015changes}.

There are various ways towards equitable data sharing that benefit of local communities.
As we have seen in the data sharing challenges section (Section~\ref{DSC}), responsible data sharing not only requires knowledge of the context of data, it also needs to center the interests and norms of communities where data comes from. Data sharing calls that are not aware of local norms, contexts, and culture, when imposed from the outside, constitute a form of Western-centrism and colonialism. For those coming from outside to participate in data sharing activities, learning ``the ways of the people'' without infiltrating their cultures need to be considered a vital part of the process.
Providing sufficient support to local researchers with the aim of facilitating local-led data sharing practices can be one of the initial steps.
 
Trust is critical during data sharing. Trust is built if those involved know each other. The reason for sharing is explicitly stated to all the stakeholders involved. Each stakeholder performs their responsibility as agreed, and all stakeholders' expectations are defined, understood, and respected. Supposing all stakeholders agree to these conditions, the likelihood of further collaboration and the possibility of fair sharing increases. It is recommended that the process of data sharing should be transparent. Also, researchers based in the continent are encouraged to build a trustworthy collaboration with funders and other stakeholders. Introducing proper and adequate legal frameworks, documentation, and support for data sharing will increase trust among stakeholders. Due to uncertainties and lack of legal frameworks, several sectors in Africa are unsure of data sharing norms and practices. They are allowed to share their data, what type of data can be shared, who, and under what conditions. Such frameworks and documentation can help clarify the benefits, expected norms, and equitable data sharing practices. Such can help build trust and encourage researchers to share their data in a manner that acknowledges their labour and protects their ideas and innovations.

Equitable data sharing is also a matter of training the next generation of data scientists.
It is imperative to include African data challenges into the data science curricula since these issues are very different from other parts of the world. Aspiring data scientists must be aware of data sources, and historical records about communities where data are collected and the harm these collected data may pose.

Many of the challenges we have discussed here -- such as data sharing in the absence of awareness of local contexts and norms as well as the issue of trust -- are daunting, and lack simple answers. Nonetheless, as we have sketched in the above sections, they are challenges that will need to be overcome to reduce community harm. The much larger and nuanced issues of data sharing, such as power asymmetry and underlying coloniality, Euro and Anglo-centrism, white supremacy, are obstacles that need deeper structural and social changes. They present challenges with no clear solutions. Understanding and acknowledging their invisible operations is a vital part of the process, and may point to how power asymmetries can be re-balanced. In this work, we have brought to light these invisible forces. While our work focuses on data access and data sharing, we note, also that many of the issues discussed may impact processes across the data generation, collection, labeling, analyses, implementation, and use pipeline. Each of these, we believe, warrants further investigation.

Last, beyond shedding light on the complex data sharing and data practice challenges in the continent, our work also uplifts research on this topic done by Africans for Africans. This practice of centering African scholarship itself challenges negative stereotypes of the continent as ``uneducated population.'' There is a vast body of work examining the rapidly evolving data practices of the African continent and its relationship with the global data ecosystem.\footnote{We share a sampling of these papers, clustered by theme, on the project page.} In the same way that data sharing challenges cannot be overcome without understanding local contexts, meaningful solutions will not be reached without deep engagement with African scholarship.

\section{Acknowledgements}
Abeba Birhane was supported, in part, by Science Foundation Ireland grant 13/RC/2094 and co-funded under the European Regional Development Fund through the Southern \& Eastern Regional Operational Programme to Lero, the Science Foundation Ireland Research Centre for Software, Ireland (www.lero.ie). The authors would like to thank our interviewees, members of the Data Science Africa initiative, workshop and conference participants at the Machine Learning for Development (ML4D) workshop, the Privacy Law Scholars Conference (PLSC), the Contested Data Academic Workshop at Data and Society, the Practical Machine learning for Developing Countries (PML4DC) workshop, as well as members of the MD4SG Data Economies and Data Governance working group for invaluable feedback throughout the development of this work. We would also like to thank Maria Tatar, Thomas Laurent, and anonymous reviewers for the same.

\balance
\bibliographystyle{ACM-Reference-Format}
\bibliography{sample-base}
\balance
\appendix

\end{document}